\documentclass[twocolumn]{aastex62}
\newcommand{\ha}{H$\alpha$}    % electron temperature
\newcommand{\hb}{H$\beta$}    % electron temperature
\newcommand{\sii}{[S~{\sc ii}]}
\newcommand{\oi}{[O~{\sc i}]}
\newcommand{\oii}{[O~{\sc ii}]}
\newcommand{\oiii}{[O~{\sc iii}]}
\newcommand{\nii}{[N~{\sc ii}]}

\newcommand{\mmm}{$\rm M^3$}

\graphicspath{{./}{figures/}}
\usepackage{amsmath}
\usepackage{CJK}
\usepackage{lineno}
\received{ xx, xx, 2022}
\revised{ xx, xx, 2022}
\accepted{{ xx, xx, 2022}}
\submitjournal{ApJL}

\shorttitle{Nebular Geometry Study}
\shortauthors{Jin et al.}

\begin{document}
\begin{CJK*}{UTF8}{gbsn}

\title{Theoretically Modelling Photoionized Regions with Fractal Geometry in Three Dimensions}

%\correspondingauthor{Yifei Jin}
%\email{Yifei.Jin@anu.edu.au}

\author[0000-0003-0401-3688]{Yifei Jin (金刈非)}
\affil{Research School for Astronomy \& Astrophysics, Australian National University, Canberra, Australia, 2611}
\affiliation{ARC Centre of Excellence for All Sky Astrophysics in 3 Dimensions (ASTRO 3D)}

\author[0000-0001-8152-3943]{Lisa J. Kewley}
\affiliation{Research School for Astronomy \& Astrophysics, Australian National University, Canberra, Australia, 2611}
\affiliation{ARC Centre of Excellence for All Sky Astrophysics in 3 Dimensions (ASTRO 3D)}

\author[0000-0002-6620-7421]{Ralph S. Sutherland}
\affiliation{Research School for Astronomy \& Astrophysics, Australian National University, Canberra, Australia, 2611}

\begin{abstract}

We create a photoionization model embedded in the turbulent ISM by using the state-of-the-art Messenger Monte-Carlo MAPPINGS~V code (M$^3$) in conjunction with the CMFGEN stellar atmosphere model.
We show that the turbulent ISM causes the inhomogeneity of electron temperature and density within the nebula.
The fluctuation in the turbulent ISM creates complex ionization structures seen in nearby nebulae. 
The inhomogeneous density distribution within the nebula creates a significant scatter on the spatially-resolved standard optical diagnostic diagrams, which cannot be represented by the spherical constant density photoionization model.
We analyze the dependence of different optical emission lines on the complexity of nebular geometry, finding that the emission-lines residing on the nebular boundary are highly sensitive to the complexity of nebular geometry, while the emission-lines produced throughout the nebula are sensitive to the density distribution of the ISM within the nebula.
Our fractal photoionization model demonstrates that a complex nebular geometry is required for accurate modeling of {H~\sc{ii}} regions and emission-line galaxies, especially for the high-redshift galaxies, where the ISM is highly turbulent based on the increasing observational evidence.

\end{abstract}

\keywords{galaxies, abundances --- 
galaxies: fundamental parameters --- 
galaxies: high-redshift}

\section{Introduction}\label{sec:intro}

Robust photoionization models are fundamental to understand galaxy evolution and chemical evolution. 
Theoretical emission-line models are required to measure the chemical abundance \citep{McGaugh-1991,Kewley-2002,Dopita-2006,Kewley-2019}, the electron density \citep{Osterbrock-1989,Kewley-2019} and the pressure of interstellar medium (ISM) \citep{Kewley-2019}. 
With these diagnostic tools, we can analyze the chemical evolution of galaxies \citep{Kobulnicky-2004,Tremonti-2004,Zahid-2011,Yuan-2013,Strom-2022}, the star formation history of galaxies \citep{Hopkins-2006,Madau-2014,Tacchella-2022}, and the power sources of galaxies across cosmic time \citep{Kewley-2013b,Rigby-2021}.

One of the most powerful applications of theoretical photoionization models has been the determination of power sources of galaxies using strong emission-lines available in the optical spectra of galaxies.
\cite{Baldwin-1981} first proposed usage of the \nii/\ha\ and \oiii/\hb\ ratios to classify the excitation sources in galaxies, which is known as the Baldwin--Phillips--Terlevich (BPT) diagram. 
The \nii/\ha\ and \oiii/\hb\ ratios excited by active galactic nuclei (AGN), shocks and planetary nebulae are larger than the ratios excited by star formation due to the difference in the hardness of ionizing radiation fields.
\cite{Veilleux-1987} derived semi-empirical classification scheme and extend the standard optical diagnostic line ratios to include the \sii/\ha\ and \oi/\ha\ ratios.

\cite{Kewley-2001} derived the first theoretical classification scheme for local galaxies by computing photoionization and shock models with the MAPPINGS III photoionization code \citep{Binette-1985,Sutherland-1993} in conjunction with the stellar population synthesis models \citep{Leitherer-1999}. 
\cite{Kewley-2013a,Kewley-2013b} further model the changes of the diagnostic line ratios as a function of cosmic time, finding the \nii/\ha\ and \oiii/\hb\ ratios are likely to be large in high-redshift galaxies due to the extreme ISM conditions in high-redshift \citep{Steidel-2014,Shapley-2005}. 

All previous models for the spectra in star-forming galaxies assume a spherical or plane parallel geometry.
However, spatially-resolved spectra of local {H~\sc{ii}} regions show the complex structures within the nebula.   
Kinematic and morphology studies reveal the filaments \citep{Gendron-Marsolais-2018,Zavala-2022} and turbulent structures \citep{Rubin-2011} within the nebula

These complex structures are not spherical or plane-parallel (Jin et al. in prep).
\cite{Snijders-2007} show that a clumpy photoionization model is required to reproduce the extremely high ionized densities and ionization parameters as found in the local starburst galaxies M82 and the Antennae.

In M82 and the Antennae, the young massive {H~{\sc ii}} regions are studied as candidates of the star forming regions in high-redshift galaxies through their dense and compact morphology \citep{Smith-2006} and the large ionization parameters \citep{Snijders-2007,Rigby-2011}.
In particular, the large ionization parameters seen in local extreme star forming regions appear to be widespread in the current set of high-redshift galaxies observed using near infrared spectroscopy \citep{Shirazi-2014,Sanders-2016,Papovich-2022}. 

In this Letter, we present the first geometrical photoionization model using over 30 chemical elements embedded in the turbulent ISM by using \mmm\, \citep[Messenger Monte-Carlo MAPPINGS~V][]{Jin-2022}, with detailed stellar atmosphere models as the input ionizing source.
We present the complex structures of the ionizing radiation field, the electron temperature, the electron density and the ionization states of different elements within the model.
We investigate how the nebular geometry affects the standard optical diagnostic emission-lines in the scope of both spatially-resolved and integrated spectroscopy.
Our results have important implications for studies of emission-line ratios in high-redshift galaxies.

\section{The method}\label{sec:meth}

\begin{figure*}
  \centering
  \includegraphics[width=5in]{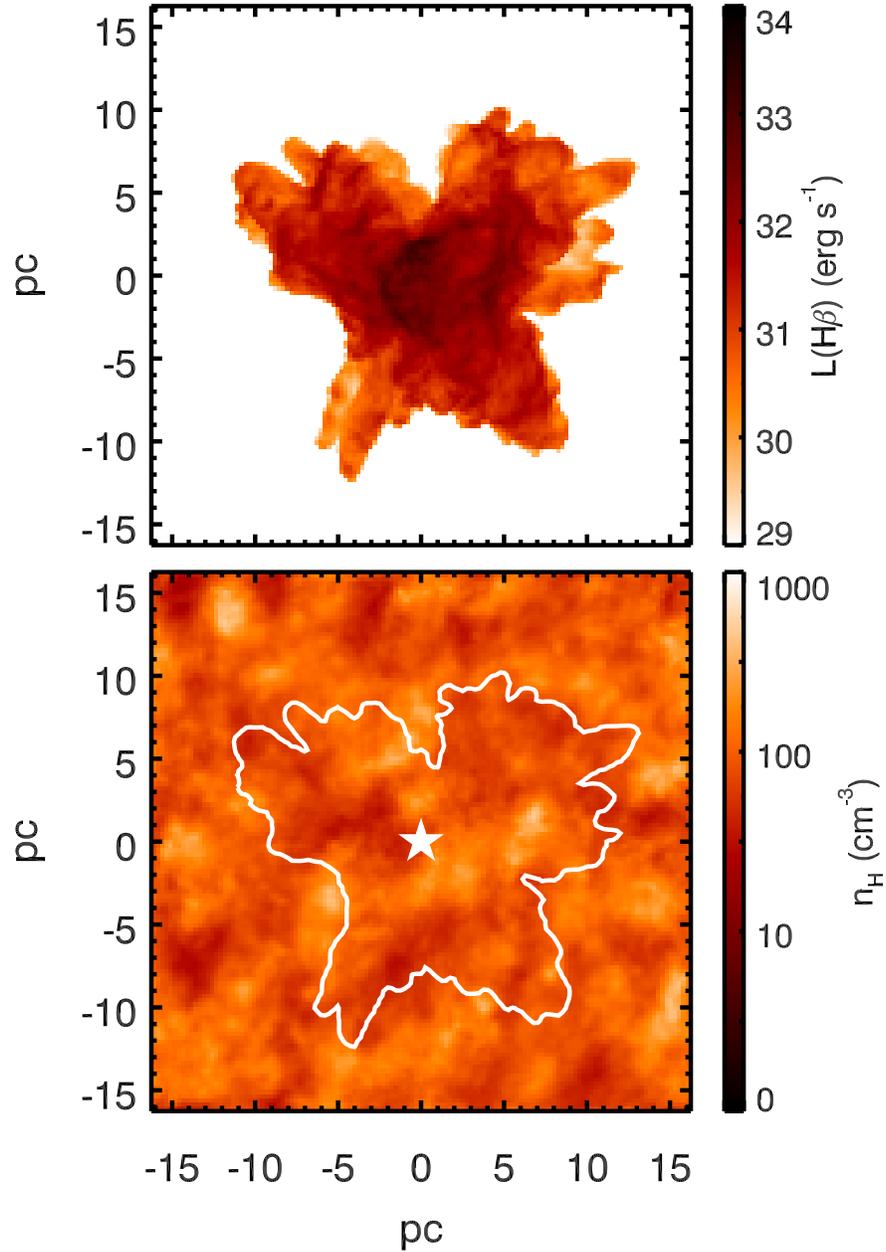}
  \caption{Upper panel: The projected \hb\ luminosity map of 3D nebula with fractal geometry. Lower panel: The projected distribution of log-normally distributed ISM. The white curve outlines the shape of our modelled nebula. The ionizing source is at the centre of the domain (indicated by the white star).}\label{fig:fig1}
\end{figure*}

\mmm\ \citep{Jin-2022} is a descendant of the MAPPINGS~V photoionization code \citep{Sutherland-1993}.
\mmm\ combines the Monte-Carlo radiative transfer technique \citep{Lucy-1999} with MAPPINGS~V \citep{Binette-1985}, which comprehensively models the microphysics of over 30 chemical elements in the ISM. 
\mmm\ is designed to create photoionization models with arbitrary geometry in three dimensions, by considering both stellar and diffuse ionizing radiation fields.

Our model is set up in a fractal ISM density field (thereafter $the$ $fractal$ $model$) to mimic the hierarchical structures of turbulence in ISM \citep{Larson-1981}.
Following \cite{Sutherland-2007}, we generate the fractal ISM density field by adopting a standard log-normal density distribution with $\mu=1.0$ and $\sigma^{2}=5.0$ and a Kolmogorov spatial structure power law.
The average density of the ISM density field is $n_H$=100~$\rm cm^{-3}$.
The static log-normal density distribution is mathematically well-constrained and shows the structure and density fluctuations similar to real turbulent molecular clouds.
The fractal ISM density field reproduces the clumps and turbulent structures seen in the real nearby HII region, like Orion \citep{Rubin-2011}.
We adopt the solar abundance based on the results from \cite{Asplund-2009}.
In this Letter, we are creating a dust-free photoionization model without dust depletion.

We select the ionizing spectrum from the CMFGEN stellar library \citep{Hillier-1998,Hillier-1999}. 
The CMFGEN library accurately describes the UV stellar spectra by using non-LTE stellar atmosphere. 
In this work, we choose an O-star with the luminosity $L$=$10^6$~$L_\sun$, the temperature $\rm T_{eff}$=40000~K and the gravity lg($g$)=4. 
The selected spectrum well describes the hard ionizing radiation from massive stars \citep{Simon-Diaz-2008}. 

Our model is set up in a cube with 129$\times$129$\times$129~spaxels (spatial pixels). 
The physical size of the entire computational domain is 43$\times$43$\times$43 pc$^3$.
The size is selected based on the size-density relation for the {H~{\sc ii}} regions in both star-forming galaxies and starburst galaxies \citep{Hunt-2009}.
A spatial resolution of 0.25~pc per spaxel guarantees that the sub-structures of the nebulae can be resolved. 

We also produce an equivalent spherical nebula model as a reference to illustrate the geometric effects on emission-line distributions, emission-line fluxes and ratios.
The spherical model is created by MAPPINGS~V with an assumption of spherical nebular geometry, in conjunction with the stellar ionizing spectrum from the CMFGEN stellar library.
Except for the geometry, the input parameters of the spherical model are the same as the input parameters of the fractal model.

\section{Results}\label{sec:result}

\subsection{The Radiation Bounded Nebula}

Figure~\ref{fig:fig1} presents the 2D map of the integrated \hb\ emission-line luminosity of the fractal model. 
The nebula has a fractal boundary consisting of pixels ionized by both stellar and diffuse photons. 
The nebula shape is determined by the surrounding ISM clumps. 
The photoionized region extends furthest in low-density regions around the ionizing source. 
The low density of the ISM in these regions allows photons to easily pass through. 
In the dense regions, the ionizing photons are absorbed by the dense clumps, leading to a nebular boundary much closer to the central ionizing source. 
Beyond the dense nebular boundary, there are regions ionized by diffuse photons, which are scattered by dense clumps. 
These regions which are ionized by the diffuse photons rather than the stellar photons are defined as the diffuse ionized regions.

Top left panel in Figure~\ref{fig:fig2} shows the local ionization parameter distribution across the middle slice of the nebula.
The local ionization parameter is defined as:
\begin{equation}
U=\frac{f_{\nu>13.6eV}}{c N_H}
\end{equation}
where $f_{\nu>13.6eV}$ is the flux of the ionizing photons above 13.6~eV through a unit area, $N_H$ is the local number density of hydrogen and $c$ is the speed of light. 
The ionization parameter indicates the capability of ionizing radiation field to ionize neutral gas.
The average dimensionless ionization parameter is log$U$=-2.8, within the range of the typical ionization parameter -3.2$<$log$U$$<$-2.7 for local {H\sc{ii}} regions \citep{Dopita-2000} and star-forming galaxies \citep{Moustakas-2010}.
The diffuse ionized regions have an ionization parameter as low as log$U$=-5 because these regions are ionized by weak diffuse ionizing photons.

The electron temperature (top central panel in Figure~\ref{fig:fig2}) ranges from 5000~K to 10000~K with average inhomogeneity of $\langle T_e \rangle$=700~K within the main body of the nebula. 
The electron temperature increases with radius on average but shows strong azimuthal asymmetry.
The positive temperature gradient is caused by the lack of coolant ions when the ionization parameter declines at large nebular radius.
The temperature is 5000~K at the nebular center and increases to 10000~K at the edge of the nebula.
The diffuse ionized regions have a low electron temperature of 2000~K because the weak ionization field can only weakly heat these regions.  
The electron temperature shows a spatial variation within the nebula. 
The high-density clumps have higher electron temperature and the low-density regions have lower electron temperature.

The electron density distribution (shown in Figure~\ref{fig:fig2} top right panel) presents a similar log-normal distribution as the neutral gas density.
The electron density spans four orders of magnitude from 0.1 $\rm cm^{-3}$ to 800 $\rm cm^{-3}$.
The average electron density is $n_e$=30~$\rm cm^{-3}$.
The inhomogeneity of electron density, which is described by the standard deviation, is $\langle n_e \rangle$=35~$\rm cm^{-3}$.

\subsection{The Ionization Structure}\label{sec:ionization}

\begin{figure*}
  \centering
  \includegraphics[width=7in]{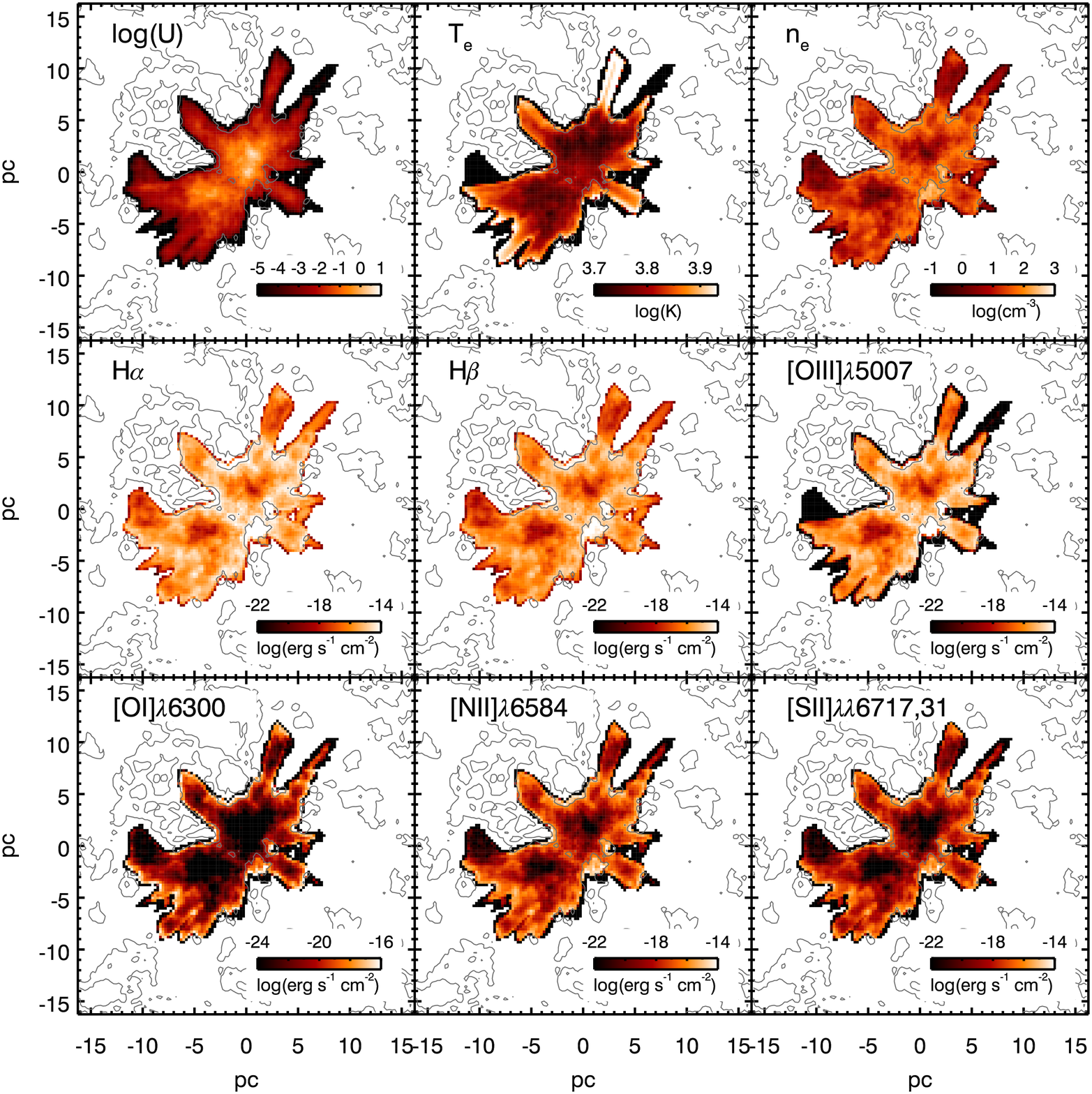}
  \caption{Maps of fundamental parameters across the Z=0~pc slice of the nebula. Grey contours indicate the distribution of the input of ISM density. 
We present maps of the dimensionless ionization parameter $U$, the electron temperature $T_{e}$, the electron density $n_{e}$ and the distribution of fluxes of the \ha , \hb , \oiii , \oi , \nii\ and \sii\ emission lines. The nebula is assumed at the distance of the Large Magellanic Cloud of 48.5kpc.}\label{fig:fig2}
\end{figure*}

Figure~\ref{fig:fig2} (lower two panels) presents the internal structure of the six diagnostic emission-lines, \ha , \hb , \oiii$\lambda$5007, \oi$\lambda$6300, \nii$\lambda$6584 and \sii$\lambda\lambda$6717,31.
These emission-lines show diverse structures based on their critical densities and ionization potentials.

The Balmer \ha\ and \hb\ lines are produced uniformly throughout the main body of the nebula, extending to the diffuse ionized region.
The variation of the \ha\ and \hb\ luminosity is smaller than one magnitude across the entire nebula.
The Balmer lines in the diffuse ionized region are fainter than the Balmer lines from the denser ionized gas.
On average, the \ha\ and \hb\ luminosity in the diffuse ionized region is 0.8 magnitude fainter than the average luminosity of the dense ionized region.

The \oiii\ emission-line is also produced throughout the main body of nebula, which is ionized by stellar photons.
In the diffuse ionized region, the \oiii\ emission is fainter than the \oiii\ emission from the nebular main body by nine magnitudes.
The extremely faint diffuse \oiii\ emission is caused by the fact that the hardness of diffuse ionization field is too low to produce $\rm O^{++}$ ions, which has the ionization potential of 35.1~eV.

\begin{figure*}
  \centering
  \includegraphics[width=7in]{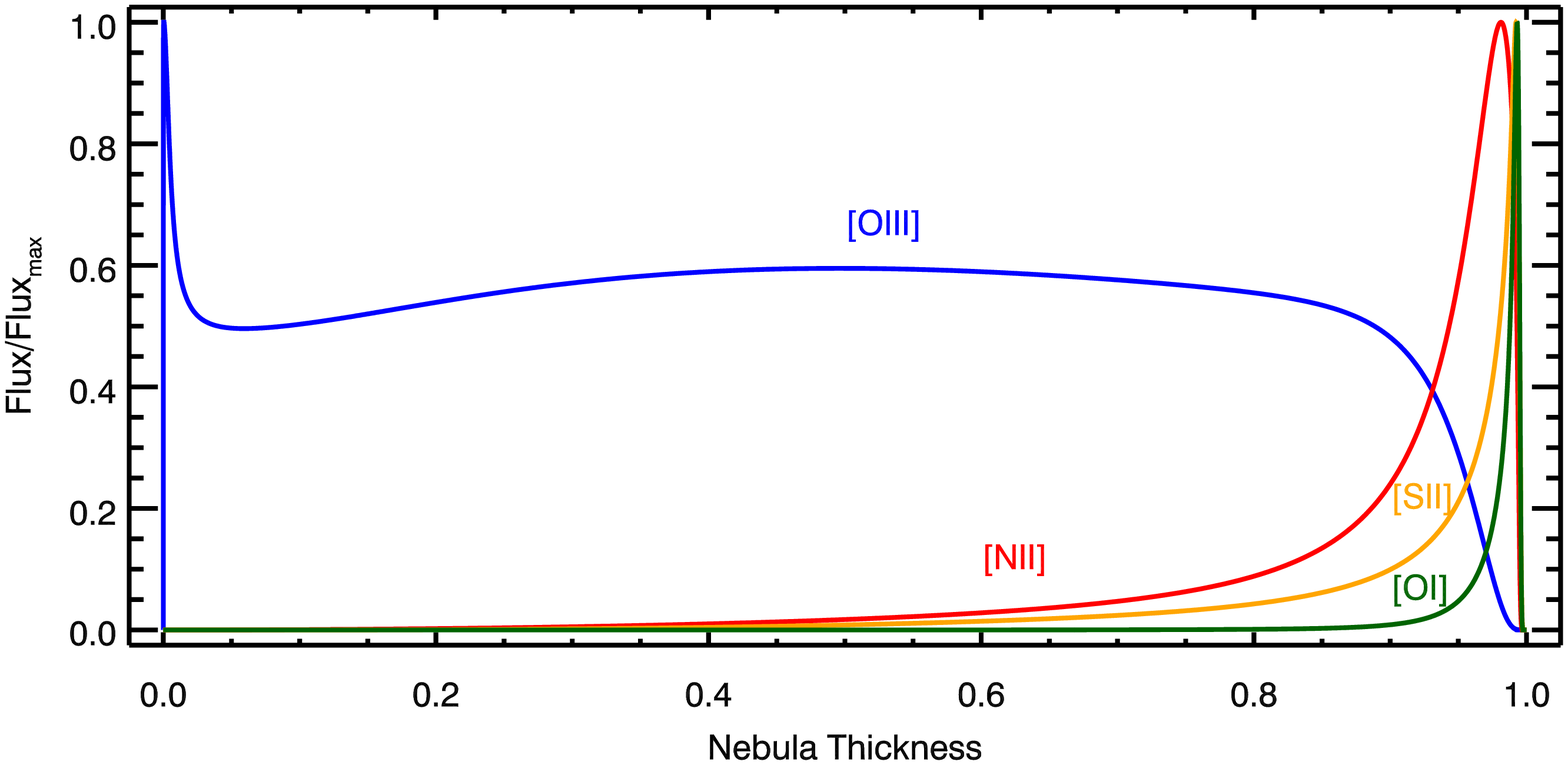}
  \caption{Spherical MAPPINGS~V model. The spherical model has the same inputs to the fractal model except for the geometry. Here, we present the radial distribution of the \oiii , \oi , \nii\ and \sii\ lines. }\label{fig:fig3}
\end{figure*}

The \oi , \nii\ and \sii\ lines are produced at the outer edge of the nebula.
As shown in Figure~\ref{fig:fig3}, 99.5\%\ of the \oi\ emission, 87.2\%\ of the \nii\ emission and 87.5\%\ of the \sii\ emission are from the outer 10\%\ radius of nebula.
In the turbulent {H\sc ii} region model, the \oi, \nii\ and \sii\ lines highlight the ``coastline'' of the nebula.
The fractal nebular boundary extends the perimeter of the outer edge of the nebula, increasing the total luminosity of \oi, \nii\ and \sii\ emission-lines compared with a standard spherical model.

The inhomogeneous density distribution of the ISM leads to the substructures of \oi, \nii\ and \sii\ lines in the fractal model.
The emission-line fluxes present a clumpy distribution within the nebula, where bright clumps trace high-density regions in ISM.
We also find an ionization bar at the center of nebula crossing from the upper-left to the lower-right of the z=0 slice.
The substructures seen in the fractal model are similar to the ionization structures seen in the Orion Nebula, where the complex ionization structures coexist with the complex density structures \citep{Rubin-2011}.

\subsection{The Spatially-Resolved Standard Optical Diagnostic Diagrams}

\begin{figure*}
  \centering
  \includegraphics[width=7in]{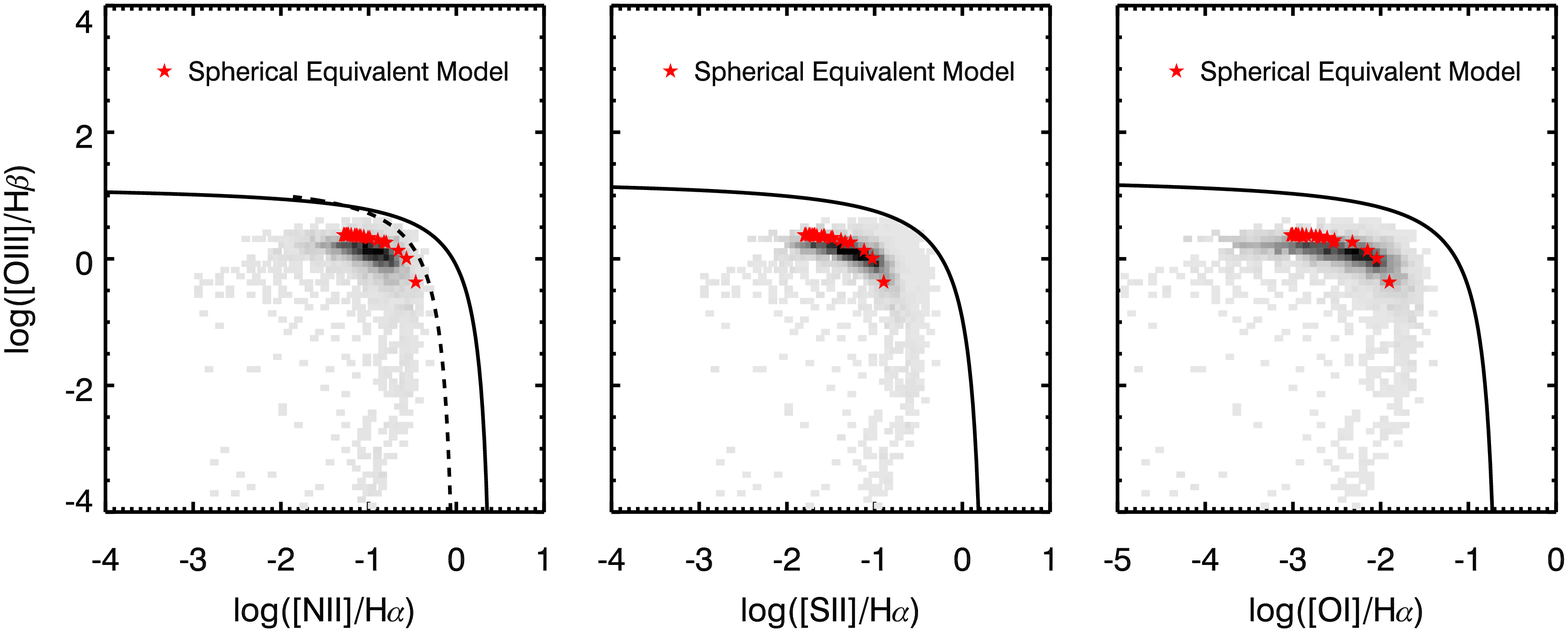}
  \caption{The comparison of diagnostic emission-line ratios from the fractal model against the spherical photoionization model. We employ the \nii/\ha\ versus \oiii/\hb\ ({\bf left column}), \sii/\ha\ versus \oiii/\hb\ ({\bf middle column}) and \oi/\ha\ versus \oiii/\hb\ ({\bf right column}) diagrams. The background grey scale indicates the number density of spaxels from the collapsed 2D map of the fractal model. The red stars are the spaxels from the imitated long-slit data of the spherical model. The solid line in each panel is the demarcation line given by \cite{Kewley-2001} to separate star-forming galaxies/H~{\sc ii} regions from AGN. The dashed line in the \nii/\ha\ versus \oiii/\hb\ panel is the demarcation line given by \cite{Kauffmann-2003} to separate star-forming galaxies/H~{\sc ii} regions from AGN.}\label{fig:fig4}
\end{figure*}

\begin{figure*}
  \centering
  \includegraphics[width=7in]{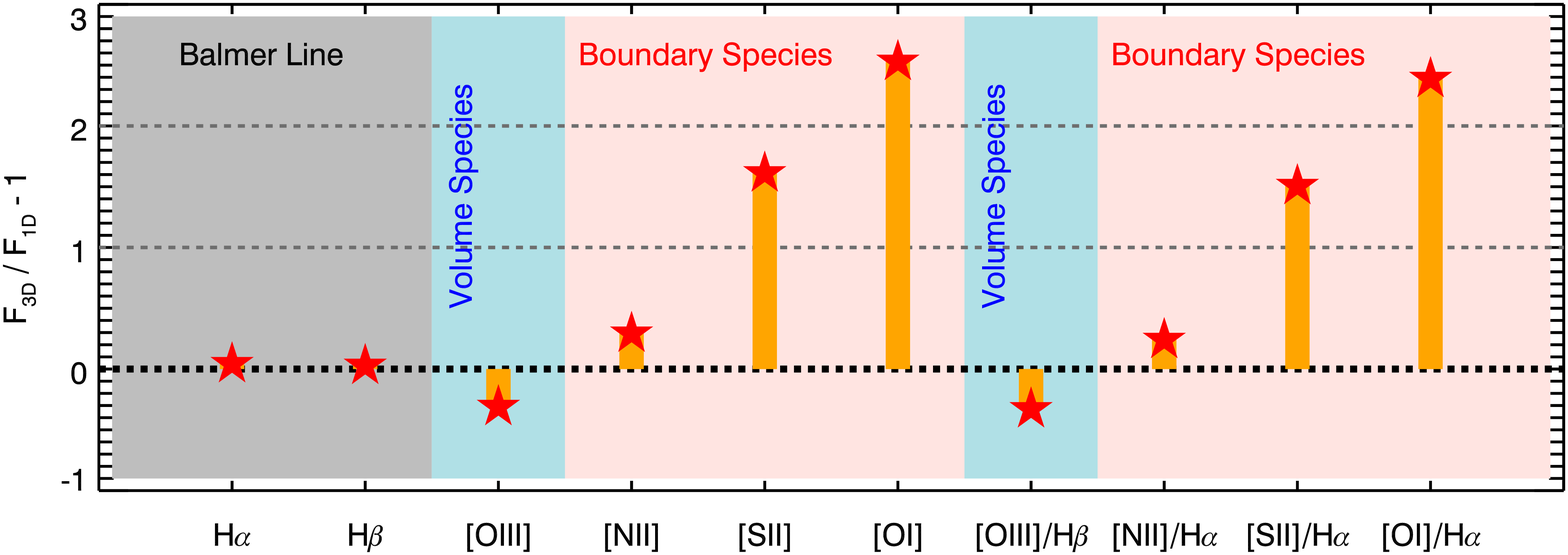}
  \caption{The comparison of integrated emission-lines fluxes and ratios between the fractal model and the spherical model. The emission-lines are classified into three categories, which are the Balmer lines ({\bf grey shadow}), the $Volume~species$ ({\bf blue shadow}) and the $Boundary~species$ ({\bf red shadow}). The black horizontal dashed line indicates no changes of the line fluxes and ratios between the fractal model and the spherical model. }\label{fig:fig5}
\end{figure*}

We investigate the \nii/\ha , \sii/\ha , \oiii/\hb , \oi/\ha\ and \oiii/\hb\ diagnostics predicted from the fractal model in comparison with the spherical model used in previous theoretical classification schemes for the optical diagnostic diagrams.
We collapse the 3D distributions of emission-lines into 2D maps to imitate spatially-resolved data.

Figure~\ref{fig:fig4} shows the number density distribution of the spaxels from our model on the optical diagnostic diagrams.
On all diagnostic diagrams, the spaxels lie on a sequence in the locus of {H~\sc{ii}} regions bounded by the demarcation between star formation and harder ionizing sources given in \cite{Kewley-2001}.  
The spaxel sequence shows a significant scatter on the diagnostic diagrams with a scatter of 0.8~dex in the log(\oiii/\hb) ratio, 0.6~dex in the log(\nii/\ha) ratio, 0.8~dex in the log(\sii/\ha) ratio and 1.2~dex in the log(\oi/\ha) ratio. 
The scatter is defined as the region that includes 16\%\ to 84\%\ of the spaxels in the emission-line ratio distributions. 
The scatter in the diagnostic line ratios is caused by the complex internal structures of the ionization parameter and the hardness of the ionizing radiation field, which vary from spaxel to spaxel, as shown in Figure~\ref{fig:fig2}.

We collapse the spherical model along the x-axis and the y-axis to produce a 1D data to imitate a long-slit observation.
Figure~\ref{fig:fig4} shows that the spatially-resolved spherical model lies on the spaxel sequence of the fractal model and has agreement with the location on the standard optical diagnostic diagrams where the spatially-resolved fractal model spaxels are concentrated.
The spherical model pixels agree with 19\%\ of the fractal model spaxels on the \nii/\ha\ versus \oiii/\hb\ diagram, 24\%\ of the fractal model spaxels on the \sii/\ha\ versus \oiii/\hb\ diagram, and 20\%\ of the fractal model spaxels on the \oii/\ha\ versus \oiii/\hb\ diagram.
However, the spatially-resolved fractal model covers a larger range of \nii/\ha , \sii/\ha , \oi/\ha\ and \oiii/\hb\ ratios than the spatially-resolved spherical model.

\subsection{Geometric Effects On Total Emission-Lines Fluxes Over A Nebula}

Figure~\ref{fig:fig5} gives the deviation between these two models between the spherical and fractal models for the total (integrated) emission-line fluxes for each line. 

The complex geometry makes only a minor change in the fluxes of the Balmer lines of 4\%\ in the \ha\ flux and 3\%\ in the \hb\ flux.
This result is expected: \cite{Kennicutt-2012} previously found that the fluxes of Case B recombination Balmer lines are solely correlated with the ionizing luminosity of the central source with limited dependence on the nebular geometry.

The nebular geometry has a pronounced influence on the forbidden lines.
We classify the forbidden lines into ``$Volume~species$'' and ``$Boundary~species$'' based on their locations within nebula (Figure~\ref{fig:fig3}). 

``$Volume~species$'' lines are the forbidden lines which are produced throughout the entire nebula and can trace the properties of the overall volume of the ISM within the nebula, \oiii\ is an example of a volume-sensitive emission-line.
In this work, the total flux of the \oiii\ line of the fractal nebula model is 31\% lower than the flux of the spherical model.
The \oiii/\hb\ ratio of the fractal model is also 33\%\ lower than the ratio of the spherical model.
The reduced \oiii\ flux is caused by the log-normal density distribution where 93\%\ of the nebular volume are occupied by the spaxels with density lower than the average density of 100~$\rm cm^{-3}$.

``$Boundary~species$'' lines are the ionization lines residing on the boundary of nebula and are sensitive to the perimeter of the nebular boundary. 
Complex nebular geometry elongates the nebular boundary, increasing the total flux of ``$Boundary~species$'' lines.
In this work, the \nii , \sii\ and \oi\ lines are the ``$Boundary~species$'' lines. 
Compared to the spherical model, the complex geometry of the fractal model increases the total emission-line flux by 29\%\ for the \nii\ line, 161\%\ for the \sii\ line and 253\%\ for the \oi\ line.
The total flux of the ``$Boundary~species$'' is positively correlated with the amount of emission-lines concentrated on the edge of nebula.
As shown in Section~\ref{sec:ionization}, 99.5\%\ of the \oi\ emission, 87.5\%\ of the \sii\ emission and 87.2\%\ of the \nii\ emission is concentrated on the boundary. 
Therefore, the \oi\ line has the largest change in total flux from the spherical to the fractal model among the three lines while the \nii\ has the least change in total flux.
Correspondingly, the line ratio is 24\%\ larger for the \nii/\ha\ ratio, 151\%\ larger for the \sii/\ha\ ratio and 239\%\ larger for the \oi/\ha\ ratio for the fractals model, relative to the 1D model.

\section{conclusions}\label{sec:conc}

We compute the first fractal-geometry photoionization model for an HII region with 30 chemical elements to estimate the electron temperature, density and emission-lines throughout the nebula in 3D.
Our fractal model shows that the density fluctuation of the turbulent ISM can cause the inhomogeneity of the electron temperature and electron density, and lead to the complex ionization structures seen in nebula.
The fractal geometry more strongly impacts emission-line fluxes of boundary elements \oi , \sii\ and \oiii .  The emission-lines of \oiii\ and the Balmer lines are only marginally affected by nebular geometry.

We find that the spatially-resolved standard optical diagnostic diagrams of the fractal model cannot be reconstructed by an equivalent spherical model, but the average emission-line ratios of these two models have agreement within 0.64~dex for {log(\oiii/\hb)} ratio, 0.16~dex for {log(\nii/\ha)} ratio, 0.13~dex for {log(\sii/\ha)} ratio and 0.23~dex for {log(\oi/\ha)} ratio.
In conjunction with the complex nebular geometry, the fluctuation of the ISM density within the fractal nebula causes a large scatter in the optical diagnostic diagrams, corresponding to different ionization parameter, density and temperature regions across an HII region.
The photoionization model with uniform density distribution and spherical geometry is insufficient to interpret the spatially-resolved observations of nearby nebula.

Geometric photoionization models are the future in interpreting emission-lines from complex ISM conditions, especially for high-redshift galaxies which have the evidently turbulent ISM.
The ISM turbulence produces the complex {H~\sc{ii}} region geometry and the density inhomogeneity within nebulae.
The perturbations from gas accretion and minor mergers in high-redshift galaxies drive the turbulence of the ISM.
Therefore, a more complex nebular geometry driven by turbulence may play a role in the larger \nii/\ha\ ratios seen in high redshift galaxies.

\vspace{25pt}

This research was conducted on the traditional lands of the Ngunnawal and Ngambri
people.
This research was supported by the Australian Research Council Centre of Excellence for All Sky Astrophysics in 3 Dimensions (ASTRO 3D), through project number CE170100013. 
YJ and LK gratefully acknowledge the support of Lisa Kewley's ARC Laureate Fellowship (FL150100113).

\clearpage

\end{CJK*}
\end{document}